\documentstyle[aps,epsf]{revtex}
\begin{document}
\title{In--out intermittency in PDE and ODE models of axisymmetric 
mean--field dynamos}
\author{
Eurico Covas${}^{1}$,
Reza Tavakol${}^{1}$,
Peter Ashwin${}^{2}$,
Andrew Tworkowski${}^{3}$ and
John M. Brooke${}^{4}$}
\address{
1.Astronomy Unit, Mathematical Sciences, Queen Mary \& Westfield
College, Mile End Road, London E1 4NS, UK
\\
2.Department of Mathematical and Computing Sciences, University of
Surrey, Guildford GU2 5XH, UK
\\
3.Mathematics Research Centre, Mathematical Sciences, Queen Mary \&
Westfield College, Mile End Road, London E1 4NS, UK
\\
4.HPC Support Group Manchester, Computing Centre, Oxford Road,
Manchester M13 9PL, UK
}
\date{\today}
\twocolumn
\maketitle
\begin{abstract}
Employing some recent results in dynamics of systems with invariant
subspaces we find evidence in both truncated and full axisymmetric mean--field 
dynamo models of a recently discovered type of intermittency,
referred to as in--out intermittency.  This is a generalised form of
on--off intermittency that can occur in systems that are not skew
products. As far as we are aware this is the first time detailed
evidence has been produced for the occurrence of a particular form of
intermittency for such deterministic PDE models and their truncations.
The specific signatures of this form of intermittency make it possible
in principle to look for such behaviour in solar and stellar
observations. Also in view of its generality, this type of intermittency
is likely to occur in other physical models with invariant subspaces.
\end{abstract}
An important feature of the Sun and stars is their variability on a wide
range of time scales \cite{gough}. In addition to the directly observed
nearly periodic magnetic solar cycles, with an average period of 22 years,
a particularly important feature of this variability is the presence of
episodes, such as the so called {\em Maunder Minimum} of the 17$^{th}$
century, during which solar activity (as deduced from sunspot numbers)
virtually vanished \cite{eddy76}. The proxy data seems to indicate that
this episode was not singular, but was preceded by numerous similar
events occurring intermittently with an intermediate time scale of the
order of $10^2$ years. There is also some evidence suggesting similar
variability in solar type stars \cite{stuiver}.

Given the absence of naturally occurring mechanisms with such time
scales in the Sun and stars, as well as evidence for the presence of
non-linear phenomena in stellar and solar magnetic activity, one of the
more plausible suggestions has been to associate this type of
variability with some form of dynamical {\it intermittency} in the
magnetohydrodynamical dynamos operating in the solar/stellar
interiors \cite{tavakol78,zeldovichetal83,spiegel85,feudel93}.

The complexity of the nonlinear partial differential equations (PDE)
modelling these regimes has essentially led to three approaches, in turn
employing:

\noindent {\bf 1.} Low dimensional models which encode the main features
of dynamo models and use, for example, normal form theory
(see e.g.\ \cite{tobias95}). These models, however, may not necessarily possess
properties which we shall argue below are generic in axisymmetric dynamos.

\noindent {\bf 2.} Low dimensional truncations 
\cite{feudel93,weiss84,ruzmaikin,weiss90}.

\noindent {\bf 3.} Direct numerical integration of the full PDE models
(e.g.\ \cite{gilman83,brandenburg89}).

Models of type 1 have produced many useful insights into the nature of
dynamo models. Models of type 2 have been shown conclusively to be
capable of producing a number of different types of intermittency,
including crisis intermittency \cite{covas97c} and type I
intermittency \cite{covas97d}. Numerical studies of models of type 3
have produced a rich set of dynamical
behaviours (see e.g. \cite{brandenburg89,tavakol95}), as well as intermittent
types of behaviour \cite{ttbbmt,brooke98,covas98a}. The problem has,
however, been how to make precise the nature of these numerically
produced dynamical modes of behaviour. This is crucial for two reasons:
firstly to be sure that these models are indeed capable of producing
intermittent behaviour and secondly to use this information to
characterise precisely their nature in order to compare their predicted
signatures with observational data.

The main difficulty has been the absence of an appropriate theoretical
framework underlying such systems. Here, we shall use recent results in
the transverse stability of attractors with invariant subspaces
\cite{covas97d,ashwin98a} to demonstrate the presence in axisymmetric
mean--field dynamo models of a recently discovered type of
intermittency referred to as {\it in--out intermittency}.

Most studies of stellar dynamos have relied on mean--field theory, which is the
approach
we adopt here. The standard mean--field dynamo equation
is given by
\begin{equation} \label{dynamo}
\frac{\partial {\bf B}}{\partial t}=
\nabla \times \left( {\bf u} \times {\bf B} + \alpha {\bf B} - \eta_t
\nabla \times {\bf B} \right),
\end{equation}
where ${\bf B}$ and ${\bf u}$ are the mean magnetic field
and mean velocity respectively and the turbulent
magnetic diffusivity $\eta_t$ and the coefficient $\alpha$
arise from the correlation of small scale
turbulent velocities and magnetic fields \cite{krause}.
In axisymmetric geometry, eq.\ (1)
is solved by splitting the magnetic field into meridional and
azimuthal components, ${\bf B} = {\bf B_{p}} + {\bf B_{\phi}}$, 
and expressing these components
in terms of scalar field functions {${\bf B_{p}} = \nabla \times A 
\hat{\phi}$, ${\bf B}_{\phi} = B \hat{\phi}$}.

In the following we shall use a finite order
truncation of the one dimensional version of equation (\ref{dynamo})
along with a time dependent form of $\alpha$, obtained by using a spectral
expansion, of the form:
\begin{eqnarray}
\frac{dA_n}{dt}&=&-n^2A_n+\frac{D}{2}(B_{n-1}+B_{n+1})+\nonumber\\
&&\sum_{m=1}^{N}\sum_{l=1}^{N}{\cal F}(n,m,l)B_mC_l,\nonumber\\
\frac{dB_n}{dt}&=&-n^2B_n+\sum_{m=1}^{N}{\cal G}(n,m)A_m, \nonumber\\
\frac{dC_n}{dt}&=&-\nu n^2 C_n
-\sum_{m=1}^{N}\sum_{l=1}^{N}{\cal H}(n,m,l)A_mB_l.\label{truncated}
\end{eqnarray}
where $A_n$, $B_n$ and $C_n$ derive from the spectral expansion of the
magnetic field ${\bf B}$ and $\alpha$ respectively, ${\cal F, H}$ and ${\cal G}$
are coefficients expressible in terms of $m,n$ and $l$, $N$ is the
truncation order, $D$ is the dynamo number and $\nu$ is the Prandtl
number (see \cite{covas97d} for details).

We first of all briefly discuss the main features necessary for the
appearance of in--out intermittency \cite{covas97d,ashwin98a} and
introduce the necessary vocabulary in the context of axisymmetric
dynamo models.

\noindent {\bf I.}
{\bf Invariant subspaces}: In addition to simplifying the resulting equations
and hence aiding the numerical integration, the presence of symmetry
forces dynamo models to possess invariant subspaces.
For example,
the truncated model (\ref{truncated}) with $N=4$ is a
12--dimensional system of ordinary differential equations (ODE) with two
6--dimensional symmetric and antisymmetric invariant subspaces given by
$M_S=\{0,$ $B_1,$ $0,$ $A_2,$ $0,$ $C_2,$ $0,$ $B_3,$ $0,$ $A_4,$ $0,$
$C_4\}$ and $M_A=\{A_1,$ $0,$ $0,$ $0,$ $B_2,$ $C_2,$ $A_3,$ $0,$ $0,$
$0,$ $B_4,$ $C_4\}$ respectively. 
Similarly the PDE model (1) possesses invariant subspaces
$M_S = \{B(\theta) =
B(-\theta)$, $A(\theta) = -A(-\theta)\}$ and $M_A =\{B(\theta) = -B(-\theta)$,
$A(\theta) = A(-\theta)\}$, where $\theta$ is the latitude.
As a result, a trajectory starting in
either subspace remains in that subspace for all times.

\noindent {\bf II.}
{\bf Non--skew product}: If the dynamics can be written as an evolution
in an invariant subspace forcing the transverse dynamics (skew product)
we cannot have proper in--out intermittency but can have on--off intermittency.
It is a generic property that we do not have a skew product.

A related feature that seems to aid the appearance of in--out
intermittency is the presence of {\it non--normal} parameters
that vary the system within the invariant subspace as well as
outside it. In the case of the truncated system (\ref{truncated}), both the
dynamo number $D$ and the Prandtl number $\nu$ are clearly non--normal
parameters, as they enter the equations for $A_n$ and $C_n$
respectively and these variables in turn are a part of the invariant
subspaces $M_S$ and $M_A$. We note that generic parameters are non--normal.

Recent studies of systems with these features 
\cite{covas97d,ashwin98a} show the
presence of a number of novel types of dynamical behaviour,
including in--out intermittency.

To characterise in--out intermittency, it is best to contrast it with
on--off intermittency \cite{ashwinplattspiegel} as both types of
intermittency occur in systems with invariant subspaces. On-off
intermittency can occur as the result of an instability of an attractor
in an invariant subspace. It manifests itself as an attractor whose
trajectories get arbitrarily close to an attractor for the system in
the invariant subspace while making occasional large deviations away
from it intermittently \cite{ashwinplattspiegel}. It can be modelled
by a biased random walk of logarithmic distance from the invariant
subspace \cite{ashwinplattspiegel}. 

We say an attractor $A$ exhibits in--out
intermittency to the invariant subspace $M_A$ (or $M_S$), if the
following are true \cite{ashwin98a}: 

\noindent {\bf 1.}
The intersection $A_0=A\cap M_A$ (or $A_0=A\cap M_S$) is not necessarily
a minimal attractor, i.e. there can be proper subsets of $A_0$ that are
attractors (for on--off intermittency $A_0$ is assumed to be minimal).
This means that there can be different
invariant sets in $A_0$ associated with attraction and repulsion transverse
to $A_0$ (hence the name in--out) and these growing and decaying phases
come about through different mechanisms within $M_A$ (or $M_S$).

\begin{figure}[!htb]
\centerline{\def\epsfsize#1#2{0.56#1}\epsffile{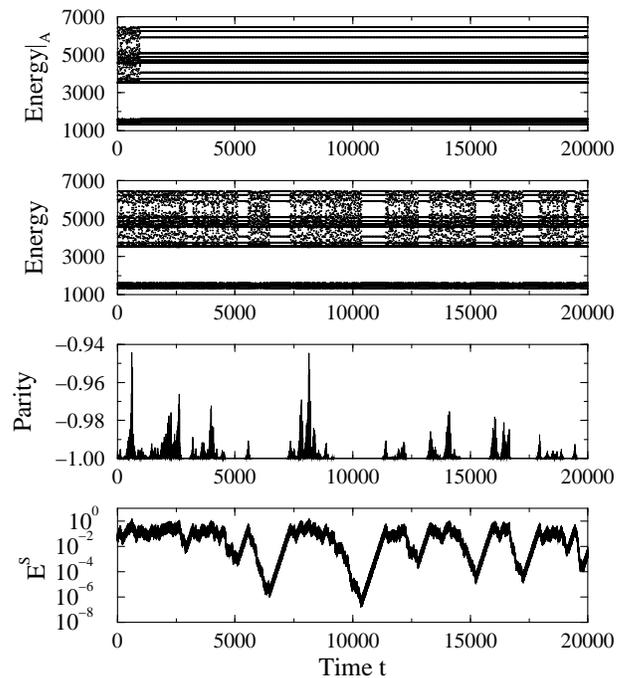}}
\caption{\label{inoutode}
In--out intermittency in the ODE model (2) with $N=4$ and parameter
values $D=177.7$ and $\nu=0.47$. The energy and parity are given by
$E=E^A+E^S$ and $P=(E^S-E^A)/E$ respectively, where $E^A$ and $E^S$ are
the antisymmetric and symmetric parts of the magnetic field energy with
respect to the rotational equator (``antisymmetric'' (P=-1) and
``symmetric'' (P=+1)). The top panel shows evolution of an initial
condition in $M_A$ and the other panels a nearby initial condition not
in $M_A$.  In these panels, we have taken a Poincar\'e section at
$A_1$=0, reducing the ODE to a map for clarity and comparison.}
\end{figure}

\noindent {\bf 2.}
The minimal attractors in the invariant subspace are not necessarily
chaotic; they can be periodic orbits or equilibria. Furthermore, the trajectory
remains close to this attractor during the moving away
or ``out'' phases, with the important
consequence that during these out phases the trajectory
can shadow for example a periodic orbit while drifting away at an exponential
rate \cite{ashwin98a} (see also \cite{brooke97}).

\noindent {\bf 3.}
The asymptotic scaling of
the distribution of laminar phases in the in--out case can have
two contributions:
\begin{equation}
P_{n}\sim \alpha n^{-3/2}e^{(-\beta n)} + \gamma e^{(-\delta n)}=
I_1+I_2,\label{scaling}
\end{equation}
where $\alpha$, $\beta$, $\gamma$ and $\delta$ are positive real constants
depending on the bias of the random walk modelling the ``in'' chain
and the probability of
leaking into the deterministic ``out'' chain \cite{ashwin98a}.
The term $I_1$ is the one
from biased on--off intermittency. The extra term $I_2$ can cause
an identifiable shoulder to develop at large $n$ which can help to
statistically distinguish this type of intermittency from on--off
intermittency.

\noindent {\bf 4.}
If the system has a skew--product structure, in--out intermittency
reduces to on--off intermittency \cite{ashwin98a}.

Because of the above considerations, we expect that in--out
intermittency will be more generally visible in such dynamo models.
Here we look at mean--field dynamo models and find in--out intermittency
in such systems.

\begin{figure}[!htb]
\centerline{\def\epsfsize#1#2{0.42#1}\epsffile{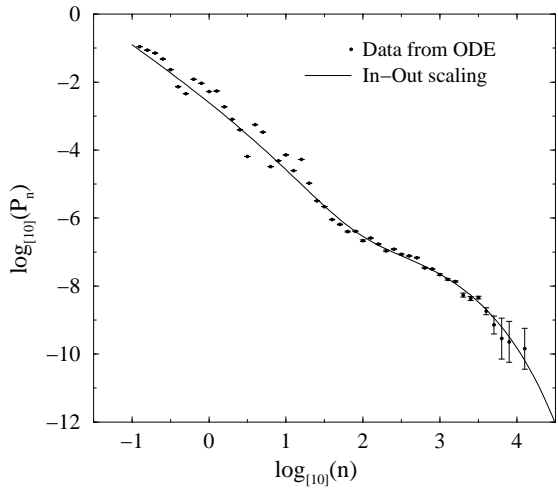}}
\caption{ \label{scalingode}
Scaling of laminar phases for the model (2) with $N=4$,
$D=177.7$ and $\nu=0.47$, where the
shoulder at large laminar phases (which is the influence of $I_2$
and a characteristic of in--out) may be discerned.}
\end{figure}

Since the ODE models are more transparent, we first look at the
truncated system (\ref{truncated}) with $N=4$.  Fig.\ \ref{inoutode}
shows an example of in--out intermittency in this system at parameter
values $D=177.7$ and $\nu=0.47$.  One can see clearly the ``periodic''
out phases (second panel), where the trajectory of the full system
shadows the periodic orbit in the antisymmetric invariant subspace $M_A$ (top
panel). Also we can clearly see the exponential growth of the amplitudes 
of the transverse variables through several orders of magnitude (lower
panel).  Furthermore, the calculated scaling of the distribution of the
laminar phases, shown in Fig.\ \ref{scalingode}, is compatible with a
curve obtained by fitting the parameters in the
scaling law (\ref{scaling}).

These signatures, namely the periodicity of the attractor of the
system restricted to the invariant submanifold, the periodic
locking and the exponential growth of the out phases 
and the compatibility with the scaling (\ref{scaling})
clearly show the occurrence of in--out intermittency for the
truncated dynamo systems. To show that this also happens in the
full PDE axisymmetric mean--field dynamo models, we first of all
recall that such systems possess the ingredients required for
the presence of in--out intermittency, namely the existence of
invariant submanifolds (symmetric or antisymmetric),
non--skew product structure
and non--normal parameters.

Guided by recent results \cite{ttbbmt}, we integrated the mean--field 
dynamo equations using the code described
in \cite{brandenburg89} and implemented by \cite{tavakol95}.

\begin{figure}[!htb]
\centerline{\def\epsfsize#1#2{0.56#1}\epsffile{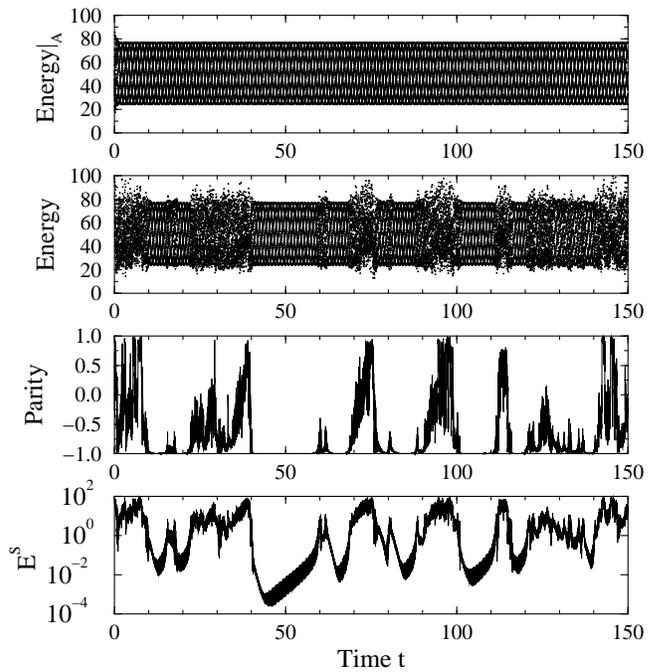}}
\caption{ \label{inoutpde1}
In--out intermittency in the axisymmetric PDE mean--field dynamo model (1).
The parameters used were $r_0=0.4$, $C_{\alpha}=1.942$,
$C_{\Omega}=-10^{5}$ ($C_{\alpha} C_{\Omega} \sim D$), $f=0.0$, with 
the usual algebraic form of $\alpha=\alpha_0/(1+{\bf
B}^2)$ (see {\protect \cite{ttbbmt}} for details of the parameters).
To enhance visually the periodic locking we time sample the 
series in the two upper panels.
}
\end{figure}

\begin{figure}[!htb]
\centerline{\def\epsfsize#1#2{0.56#1}\epsffile{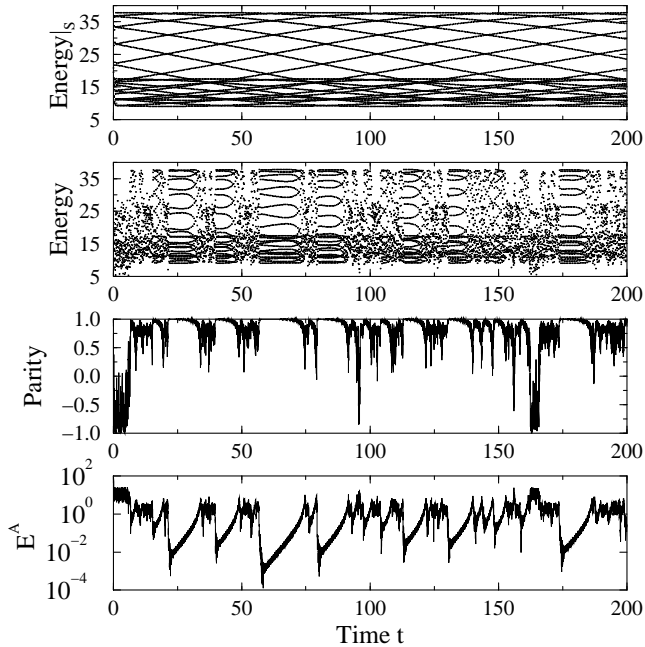}}
\caption{ \label{inoutpde2}
In--out intermittency in the axisymmetric PDE mean--field dynamo model (1).
The parameters used were $r_0=0.4$,
$C_{\alpha}=1.5$, $C_{\Omega}=-10^{5}$,
$f=0.7$, together with an algebraic form of $\alpha$ due to Kitchatinov
{\protect \cite{kitchatinov}}.
The two upper panels are shown as in Fig.\ \ref{inoutpde1}.
}
\end{figure}

Fig.\ \ref{inoutpde1} and Fig.\ \ref{inoutpde2} give examples of
in--out intermittency in these PDE models with different algebraic forms of
$\alpha$. As can be seen this behaviour can occur
with the invariant submanifold being either antisymmetric (Fig.\
\ref{inoutpde1}) or symmetric (Fig.\ \ref{inoutpde2}). We have also
obtained similar results using a dynamic form of $\alpha$ of a
similar type to that used in the truncated models. 
In addition to the presence of periodic behaviour in the system
restricted to the invariant submanifold (top panel), 
these figures clearly show the presence of
locking during the out phases (second panel) 
with an exponential growth of the energy 
of the transverse modes
(bottom panel).
This behaviour mirrors very closely 
the truncated model shown in Fig.\ \ref{inoutode} as well as that
expected to occur from the theory \cite{ashwin98a}.
Preliminary results show that the scaling for the PDE model is
also compatible with (\ref{scaling}), but a clear verification requires a much
longer integration.  We shall return to a more detailed
study of this issue in the future. The presence of the main
features necessary for the occurrence of in--out intermittency in these
models, as well as its presence in related
truncated models, constitutes strong
evidence for the occurrence of in--out intermittency in these PDE
models.

We note that similar types of behaviour have also been
seen in mean--field dynamo models with different
topologies \cite{brooke98,brooke} thus lending further support to
the suggestion that such behaviour is likely to occur in other settings.

In summary, we have found
strong evidence for the occurrence of in--out
intermittency in both truncated and PDE axisymmetric mean--field dynamo
models.
Given the specific signatures of such intermittent regimes,
this makes it in principle possible to test the presence of such
behaviour in solar and stellar observations.

We also note that since dynamical systems are generically
non--skew product with
non--normal parameters,
we expect this type of
intermittency to be present in models of other physical systems with
symmetry as well as their truncations.
\vspace{0.2cm}

We thank Axel Brandenburg and David Moss for helpful conversations.  EC
is supported by grant BD/5708/95 -- PRAXIS XXI, JNICT.  PA is partially
supported by a Nuffield ``Newly appointed science lecturer'' grant. RT
benefited from PPARC UK Grant No. L39094. This research also benefited
from the EC Human Capital and Mobility (Networks) grant ``Late type
stars: activity, magnetism, turbulence'' No. ERBCHRXCT940483.

\end{document}